\documentclass[onecolumn,pra,superscriptaddress,10pt]{revtex4}

\usepackage{graphicx}
\usepackage{dcolumn}
\usepackage{bm}
\usepackage{amssymb}
\usepackage{amsmath}
\usepackage{extarrows}

\begin{document}

\title{Simulation of the many-body dynamical quantum Hall effect in an optical lattice}

\author{Dan-Wei Zhang}
\email{zdanwei@126.com}\affiliation{Guangdong Provincial Key Laboratory of Quantum Engineering and Quantum Materials,
SPTE, South China Normal University, Guangzhou 510006, China}
\affiliation{Department of Physics and Center of Theoretical and Computational
Physics, The University of Hong Kong, Pokfulam Road, Hong Kong, China}

\author{Xu-Chen Yang}
\affiliation{Department of Physics and Materials Science, City University of Hong Kong, Kowloon, Hong Kong SAR, China}

\begin{abstract}
We propose an experimental scheme to simulate the many-body dynamical
quantum Hall effect with ultra-cold bosonic atoms in a one-dimensional optical lattice. We first show
that the required model Hamiltonian of a spin-$1/2$ Heisenberg chain with an effective magnetic field and tunable parameters
can be realized in this system. For dynamical response to ramping the external fields, the quantized plateaus emerge in the Berry curvature of the interacting atomic spin chain as a function of the effective spin-exchange interaction. The quantization of this response in the parameter space with the interaction-induced topological transition characterizes the many-body dynamical quantum Hall effect. Furthermore, we demonstrate that this phenomenon can be observed in practical cold-atom experiments with numerical simulations.
\end{abstract}



\maketitle

\section{Introduction}

The Berry curvature and its associated Berry phase \cite{Berry} have numerous important applications in physics, particularly in
quantum computation \cite{Zanardi,Sjoqvist,Duan2001,Zhu2002,Zhu2003,Xue1,Xue2,Xue3} and modern condensed matter physics
\cite{Xiao,Klitzing1980,Tsui,Laughlin,Hasan,Qi}. A prime example is the celebrated integer quantum Hall effect (QHE), which is experimentally characterized by the quantized non-diagonal conductivity with the form of $n e^2/h$. The integer $n$ here is a topological invariant, which can be formulated as the integral of the Berry curvature over the momentum space and is called as the Chern number \cite{Thouless1982,Niu1985}. Usually, the applications and experimental measurements of the Berry phase rely on the existence of free or nearly free quasi-particles that can independently
interfere and affect transport properties \cite{Zanardi,Sjoqvist,Duan2001,Zhu2002,Zhu2003,Xue1,Xue2,Xue3,Xiao,Klitzing1980,Tsui,Laughlin,Hasan,Qi,Thouless1982,Niu1985}.

Recently, it was theoretically revealed that the Berry curvature and hence the Berry phase in
generic systems (single particle and many-body systems) can be observed from a non-adiabatic response on
physical observables to the rate of change of an external
parameter, without involving interference or transport measurements \cite{Gritsev2012,Avron2011}. This phenomenon can be
interpreted as a dynamical QHE in a parameter space, i.e., the quantization of the dynamical response \cite{Gritsev2012}. In this general relation, the conventional QHE can be considered as a particular example if one views the electric field as a rate of change of the vector
potential. This work reveals deep connections between quantum dynamics and equilibrium geometric properties, and provides a new avenue to explore the QHE in the parameter space and the Berry phase effects in many-body systems \cite{Gritsev2012}. The proposal of measuring the Berry curvature was recently realized with a single superconducting qubit and two coupled superconducting qubits \cite{Schroer2014,Roushan2014}. The further simulation of the dynamical QHE with a superconducting qubit array was also proposed \cite{Yang}. However, it is extremely challenging to experimentally construct a fully-connected network of superconducting qubits with tunable inter-qubit coupling and long coherence time \cite{Schroer2014,Roushan2014,Yang,Tan,Martinis}, which may prevent the study of interaction-induced topological transition in the many-body dynamical QHE in these superconducting-qubit systems.

On the other hand, ultra-cold atoms in optical lattices provide an ideal platform for quantum simulation \cite{Nori1,Nori2} of various important models and the related effects in condensed matter physics \cite{Lewenstein,Bloch,Dalibard,Goldman,ZhangFP}. For instance, these systems have been used to realize the Bose-Hubbard model and observe the quantum phase transition between the superfluid and Mott insulator phases \cite{BH1989,BH,BHExp}, and to mimic quantum spin models and observe different magnetic phases with non-trivial dynamics \cite{Simon,Fukuhara1,Fukuhara2,Fukuhara3,Gross}. Based on the Raman transition method, recent theoretical and experimental advances have been put forward to create effective spin-orbit coupling and gauge potentials for ultra-cold atoms, and thus the related topological phases and relativistic quantum effects can be simulated \cite{Dalibard,Goldman,ZhangFP}. For these cold atom systems acting as powerful quantum simulators \cite{Nori1,Nori2}, a realistic experimental scheme for simulation of the many-body QHE is still badly needed and would be of great value.

Here, we propose an experimental scheme to simulate the many-body dynamical QHE with ultra-cold bosonic atoms in a one-dimensional optical lattice based on a state-of-the-art technique. We first show that the required model Hamiltonian of a spin-$1/2$ Heisenberg chain with an effective magnetic field and tunable parameters can be realized in this cold atom system, and then introduce the procedure of implementing the dynamical QHE. For the dynamical response to ramping the external fields, the quantized plateaus emerge in the Berry curvature of the interacting atomic spin chain as a function of the effective spin-exchange interaction. The quantization of this response in the parameter space with the interaction-induced topological transition characterizes the many-body dynamical QHE. By numerical simulations, we also demonstrate that this phenomenon can be observed under practical conditions in typical cold-atom experiments. The realization of the many-body dynamical QHE in the optical lattice may extend the current intensive study of topological properties in cold atom systems \cite{Dalibard,Goldman,ZhangFP}, and pave the way to explore intriguing topological responses with interactions.

The paper is organized as follows. Section II introduces
the realization of the model Hamiltonian of an interacting spin chain in an optical lattice. In Sec. III, we present the implementation procedure and the observation of the many-body dynamical QHE in this cold atom system with numerical simulations. In Sec. IV, some possible concerns in practical experiment are briefly discussed and finally a short conclusion is presented.

\section{Realizing the model Hamiltonian}

It has been demonstrated that the many-body dynamical QHE can emerge in the model Hamiltonian of a spin-$1/2$ ferromagnetic
spin chain with an external magnetic field, which takes the form as \cite{Gritsev2012}
\begin{equation}
\label{model}
\mathcal{H} = J\sum\limits_{j = 1}^{N - 1} {{{\vec \sigma
}_j} \cdot {{\vec \sigma }_{j +
  1}}}  +\sum\limits_{j = 1}^N {\vec h \cdot {{\vec
\sigma }_j}},
\end{equation}
where $\vec \sigma_j\equiv(\sigma_j^x,\sigma_j^y,\sigma_j^z)$ stands for Pauli matrices,
$J>0$ is the ferromagnetic spin-exchange interaction between the nearest-neighbor spins,
$\vec h\equiv(h_x,h_y,h_z)$ is the magnetic field, and $N$ is the size of the spin chain.
In the following, we will show that this model Hamiltonian with tunable parameters can be realized with ultra-cold atoms in an optical lattice and the related dynamical QHE can be observed in this system.

We start with an atomic gas of (pseudo) spin-$1/2$ interacting ultra-cold
bosons loaded in a one-dimensional spin-independent optical lattice. The spin
degree of freedom is encoded by two hyperfine states
$|\uparrow\rangle$ and $|\downarrow\rangle$. Consider the two
hyperfine states are coherently coupled via a two-photon Raman transition with a large detuning $\Delta_d$ and two Raman lasers denoted by their Rabi frequencies $\Omega_{1,2}$, as shown in Fig. 1(a). Here we assume that $\Omega_{1,2}$ are real and constants along the $x$ axis. The single-particle Hamiltonian of this system can be written as
\begin{equation}
\label{NISTHam}
\hat{H}_0=\frac{\hat{p}^2}{2m}+\delta\sigma_z+\Omega_x\sigma_x+V_0\sin^2(k_0x),
\end{equation}
where $m$ is the atomic mass, $\hat{p}$ is the momentum operator,
$2\delta$ is the resulting effective Zeeman splitting, $\Omega_x=\Omega_1\Omega_2/\Delta_d$ is the resulting effective Rabi frequency,
$V_0$ and $k_0$ respectively denote the strength and the wave number of the lattice potential.

\begin{figure}[tbp]
\includegraphics[width=16.0cm]{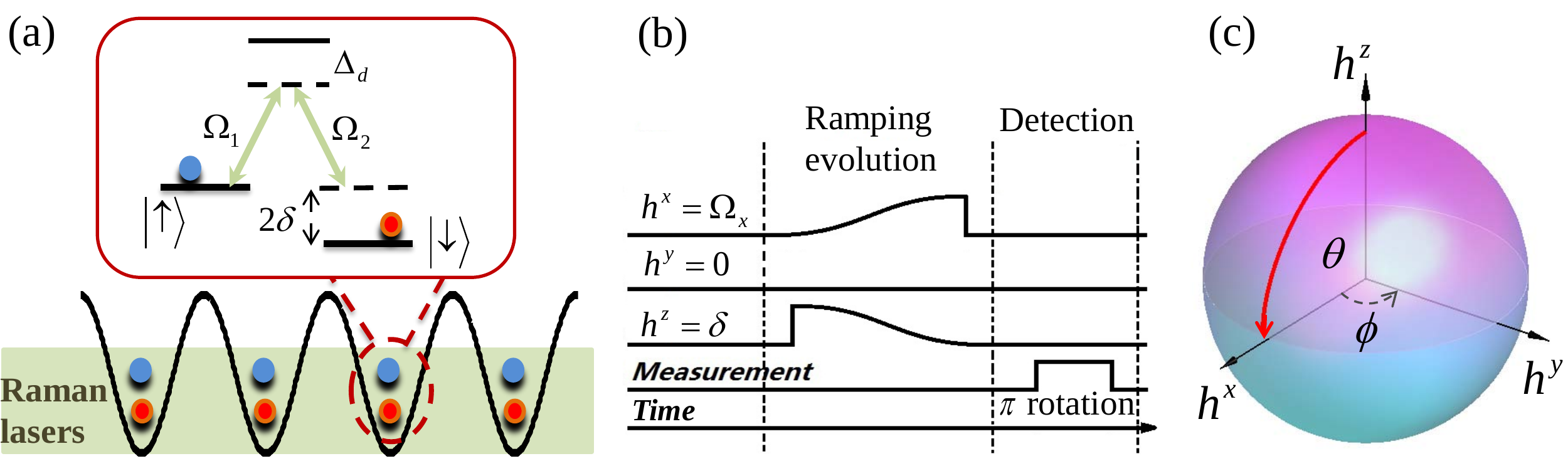}
\caption{(Color online) The cold atom system and the procedure for simulating
the dynamical QHE. (a) The schematic diagram of the system for realizing the model Hamiltonian (\ref{model}). Two-component cold bosonic atoms in a one-dimensional optical lattice are subjected to two Raman lasers denoted by their Rabi frequencies $\Omega_{1,2}$ (here $\Omega_{1,2}$ are real and constants along the $x$ axis) for the two-photon coupling with a large detuning $\Delta_d$ (i.e., $\Delta_d\gg\Omega_{1,2},\delta$) and an effective Zeeman splitting $2\delta$, resulting in the effective Rabi frequency $\Omega_x=\Omega_1\Omega_2/\Delta_d$. (b) The sequence of the ramp parameters and measurement. The atomic spin chain undergoes a non-adiabatic evolution during the ramp with the effective magnetic field followed by Eq. (\ref{field}). The magnetization of the final state is then measured. (c) The spherical diagram of the effective magnetic field, with the red solid curve representing the ramp path.}
\end{figure}

In the second quantization formalism, we can rewrite the single-particle Hamiltonian as
\begin{equation}
\mathcal{H}_0=\int dx~\Psi^{\dag}(x)\hat{H}_0\Psi(x).
\end{equation}
where $\Psi^{\dag}$ ($\Psi$) is the two-component field operator with
$\Psi^{\dag}=\left[\sum_{j}\hat{a}^{\dag}_{{j},\uparrow}w^*_{{j},\uparrow}(x),
\sum_{j}\hat{a}^{\dag}_{{j},\downarrow}w^*_{{j},\downarrow}(x)\right]$.
Here $w_{{j},\sigma}(x)$ represents the Wannier functions localized at site
$j$, and $\sigma=\uparrow,\downarrow$ denotes the atomic
pseudospin. In the tight-binding approximation, the summation can be restricted to nearest neighbors only, thus we can obtain
%
\begin{equation}
\mathcal{H}_0=-\sum_{j,\sigma}\left(t_{\sigma}\hat{a}^{\dag}_{{j},\sigma}\hat{a}_{{j}+1,\sigma}+\text
{H.c.}\right)+\Omega_x\sum_{j}\left(\hat{a}^{\dag}_{{j},\uparrow}\hat{a}_{{j},\downarrow}+\hat{a}^{\dag}_{{j},\downarrow}\hat{a}_{{j},\uparrow}\right)
+\delta\sum_{j}\left(\hat{a}^{\dag}_{{j},\uparrow}\hat{a}_{{j},\uparrow}-\hat{a}^{\dag}_{{j},\downarrow}\hat{a}_{{j},\downarrow}\right),
\end{equation}
where $t_{\sigma}=-\int  w^*_{{j,\sigma}}(x)\left[\frac{\hat{p}^2}{2m}+V_0\sin^2(k_0x)\right]w_{{j+1},\sigma}(x)dx$ is the tunneling amplitude.
Adding interaction, the total Hamiltonian of the system is given by $\mathcal{H}=\mathcal{H}_0+V_{\text{int}}$, which describes an extended Bose-Hubbard chain of ultra-cold atoms, with the term $V_{\text{int}}$ describing interatomic interactions as \cite{BH}
\begin{equation}
V_{\text{int}}=\frac{1}{2}\sum_{{j},\sigma\sigma'}U_{\sigma\sigma'}\hat{a}^{\dag}_{{j},\sigma}\hat{a}^{\dag}_{{j},\sigma'}\hat{a}_{{j},\sigma'}\hat{a}_{{j},\sigma},
\end{equation}
where $U_{\sigma\sigma'}=\frac{4\pi\hbar^2 a_{\sigma\sigma'}}{m}\int  |w_{{j,\sigma}}(x)|^2|w_{{j},\sigma'}(x)|^2dx$ with $a_{\sigma\sigma'}$ the scattering length between spins $\sigma$ and $\sigma'$.

We consider the system with unit filling (one atom per lattice) in the deep Mott insulator regime, i.e.
$|U_{\sigma\sigma'}|\gg t_{\sigma},|\delta|,|\Omega_x|$. In this parameter regime, the
atoms are localized in individual lattices and the near-neighbor
tunneling can be treated as perturbations. It is convenient to write
the lattice effective Hamiltonian in terms of isospin operators with
$\sigma_{j}^x=(\hat{a}^{\dag}_{{j},\uparrow}\hat{a}_{{j},\downarrow}+\hat{a}^{\dag}_{{j},\downarrow}\hat{a}_{{j},\uparrow})/2$,
$\sigma_{j}^y=-i(\hat{a}^{\dag}_{{j},\uparrow}\hat{a}_{{j},\downarrow}-\hat{a}^{\dag}_{{j},\downarrow}\hat{a}_{{j},\uparrow})/2$
and
$\sigma_{j}^z=(\hat{a}^{\dag}_{{j},\uparrow}\hat{a}_{{j},\uparrow}-\hat{a}^{\dag}_{{j},\downarrow}\hat{a}_{{j},\downarrow})/2$.
Up to the second order of perturbation \cite{PRL2003,duan,zhang2013,Shan}, the resulting effective spin
Hamiltonian of the system with $N$ lattice sites (spins) is
\begin{eqnarray}
\label{SpinHam}
\mathcal{H}_{S}=\sum_{j=1}^{N-1}\left(J_x\sigma_{j}^{x}\sigma_{{j+1}}^{x}+J_y\sigma_{j}^{y}\sigma_{{j+1}}^{y}+J_z
\sigma_{j}^{z}\sigma_{{j+1}}^{z}\right)+\sum_{j=1}^{N}\vec{h}\cdot
\vec{\sigma}_j.
\end{eqnarray}
Here $J_x=J_y\equiv J=-4t_{\uparrow}t_{\downarrow}/U_{\uparrow\downarrow}$ and
$J_z=-4t_{\uparrow}^2/U_{\uparrow\uparrow}-4t_{\downarrow}^2/U_{\downarrow\downarrow}+2(t_{\uparrow}^2+t_{\downarrow}^2)/U_{\uparrow\downarrow}$
are the spin-exchange interactions, and $\vec{h}=(\Omega_x,0,\delta)$ acts like an effective magnetic field on the spin chain. In the harmonic
approximation for the ground-state Wannier function \cite{BH}, the tunneling amplitude and the interaction energy can
be derived as $t_{\sigma}=t\approx(4/\sqrt{\pi})V_0^{3/4}E_R^{1/4}\exp(-2\sqrt{V_0/E_R})$ and $U_{\sigma\sigma'}\approx(8/\pi)^{1/2}k_0a_{\sigma\sigma'}(E_RV_0)^{1/4}$, where $E_R=\hbar^2k_0^2/2m$ is the recoil energy. From these expressions, one can observe that $t$ and $U_{\sigma\sigma'}$ are tunable by controlling the strength of the optical lattice, and $U_{\sigma\sigma'}$ can also be tuned from positive to negative within a broad range by changing the scattering lengths $a_{\sigma\sigma'}$ through Feshbach resonances \cite{Chin}. In our case, one focus on the parameter regime with $U_{\sigma\sigma'}<0$ and $t_{\sigma}>0$, which gives rise to a ferromagnetic spin chain \cite{duan}. In addition, the Zeeman splitting $\delta$ and the Rabi frequency $\Omega_x$ can be independently tuned. For example, one can use a real magnetic field for fixed laser frequencies or vary the Raman laser frequencies \cite {Gross} to tune $\delta$, and change the intensities of the Raman lasers to tune $\Omega_x$. Thus one can realize the interacting spin model described by Hamiltonian (\ref{SpinHam}) with tunable parameters for implementation of the many-body dynamical QHE with ultra-cold atoms in the optical lattice \cite{Gritsev2012}.

Before preceding, we estimate the typical energy scales in practical experiments. We consider $^{87}$Rb atoms and the two hyperfine states for the effective spin index as $|\uparrow\rangle=|F=1,m_F=-1\rangle$ and $|\downarrow\rangle=|F=1,m_F=0\rangle$ \cite{SOCBEC}. The interaction energies can be chosen as $U_{\uparrow\uparrow}=U_{\downarrow\downarrow}\equiv U\approx 1.01U_{\uparrow\downarrow}$ from the respective scattering lengths with negative sign \cite{SOCBEC,ScatLeng}. Thus the $z$-direction spin interaction $J_z=0.98J\approx J$, which indicates that the spin interactions in this atomic spin chain can be almost isotropic. For $^{87}$Rb atoms in an optical lattice with the lattice spacing $\pi/k_0=532$ nm and the potential depth ranging from $V_0=8E_R$ to $11E_R$ \cite{Fukuhara1,Fukuhara2,Fukuhara3}, the typical tunneling rate is from $t/\hbar\approx0.7$ kHz to $0.2$ kHz, and the interaction energy can be $|U|=-U\approx7$ kHz, yielding the spin exchange rate from $J/\hbar\approx280$ Hz to $23$ Hz. In this case, one can choose $|\Omega_x|$ and $|\delta|$ from zero to several hundreds of Hz, such that the system is still in the Mott regime in the presence of Raman lasers and a Zeeman field. In the next section, we will see that the many-body dynamical QHE can be observed within this parameter regime.

\section{Simulating the dynamical quantum Hall effect}

The topological properties of this cold atom system can be characterized by the Berry curvature. As proposed in Ref. \cite{Gritsev2012}, the dynamical QHE can be simulated in this spin chain model by observing the response of the system to ramping the effective magnetic field $\vec{h}\equiv(h^x,h^y,h^z)$. To demonstrate the dynamical QHE in this system, we consider the ramp progress with the following parameterized form
\begin{equation}
\label{field}
\begin{split}
h^x\left( t \right) &= \Omega_x(t)={h}\sin \theta  \cos  \phi  ,\\
h^y\left( t \right) &= {h}\sin  \theta \sin  \phi  ,\\
h^z\left( t \right) &= \delta(t)={h}\cos \theta.
\end{split}
\end{equation}
Here the field strength $h$ is tunable but fixed during each ramp progress, the phase parameter $\phi(t)=0$ in this system due to the absence of $h^y$ term which requires another Raman process, and the mixing angle $\theta(t)$ is used as the ramp parameter for observing the dynamical QHE.

\begin{figure}[tbp]
\includegraphics[width=16.0cm]{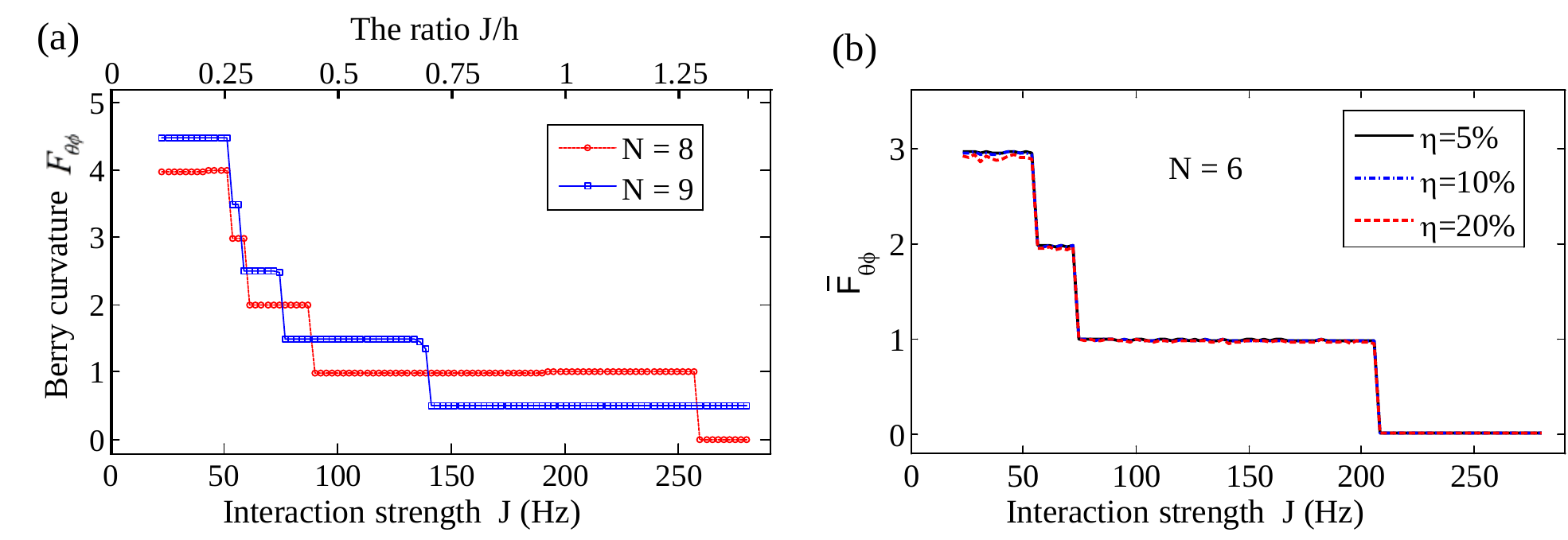}
\caption{(Color online) (a) The Berry curvature $F_{\theta\phi}$ as a function of the interaction strength $J$ for the spin systems with $N=8,9$. The ratio between the interaction strength and the filed strength $J/h$ is also shown. (b) The averaged Berry curvature $\bar{F}_{\theta\phi}$ for the spin system with $N=6$ as a function of $J$, with the local fluctuations $\tilde{J}_j=\alpha_1 J$ and $\tilde{h}_j=\alpha_2 h$ and the sampling number $N_{\alpha}=50$. Other parameters in (a) and (b) are $h=200$ Hz and $t_{\text{ramp}}= 0.01$ s. }
\end{figure}

Following the general scheme proposed in Ref. \cite{Gritsev2012}, in our system, we can prepare this ultra-cold atomic spin chain initially in the ground state with $\theta(t=0)=0$, and then ramp the system with fixed $\phi(t)=0$ to undergo a quasi-adiabatic evolution by varying the mixing angle
$\theta(t)=v^2t^2/2\pi$ for a ramp time $t_{\text{ramp}}=\pi/v$,
where $v$ denotes the ramp velocity. The condition of fixing $\phi=0$ is spontaneously satisfied in this system without additional operations in experiments. At the end of such a ramp with $\theta(t=t_{\text{ramp}})=\pi/2$ and the velocity of the $\theta$-component of the magnetic field
$v_\theta(t=t_{\text{ramp}})=\dot{\theta}(t=t_{\text{ramp}})=v$, we can measure the Berry
curvature of the interacting spin chain from its magnetization \cite{Gritsev2012}. The three components of
the effective magnetic field during the whole evolution process are depicted in Fig. 1(b). The ramp of this system can be easily achieved by independently varying the strength of the Zeeman field and the intensities of the Raman lasers in time \cite{SOCBEC}. This choice of ramping field
guarantees that the angular velocity is turned on smoothly and the system will not be excited at the beginning \cite{Gritsev2012}.

The response of the generalized force along the latitude direction for the spin Hamiltonian is measured at $t=t_{\text{ramp}}$. At this point of the measurement the generalized force $M_\theta$ is along the $y$-axis since $\theta=\pi/2$, whereas the quench velocity is along the longitude direction. Then $M_\theta$ is given by
 \begin{equation}
 \label{response}
 M_\theta = -\langle \partial_{\phi} \mathcal{H}_S \rangle \mid_{\phi=0,t=\pi/v} = -h\sum_{j=1}^{N}\langle\sigma_j^y\rangle.
 \end{equation}
From Eq. (\ref{response}), one can observe that the generalized force at the end of the ramping process is expressed by the spin magnetization along the $y$ axis, which can be measured in cold atom systems \cite{Fukuhara1,Fukuhara2,Fukuhara3}. Within the linear response approximation, it has been shown that the Berry curvature of the system $F_{\theta \phi}$ is connected to the response of the generalized force with the simple relationship \cite{Gritsev2012,Avron2011}
 \begin{equation}
 \label{force}
   {F_{\theta \phi }} = -\frac{M_\theta}{{{v_\theta }}} = \frac{h}{v}\sum\limits_{j = 1}^N {\left\langle {\sigma _j^y} \right\rangle }.
 \end{equation}
In other words, the Berry curvature reveals itself from an effective force. To measure the magnetization $\sum_j\left\langle {\sigma _j^y}
\right\rangle$ in cold-atom experiments, one can perform a global $\pi/2$ rotation on the spin chain to transform the measurement in the transverse spin basis to $\sum_j\left\langle {\sigma _j^z} \right\rangle$, and then probe this magnetization by the \textit{in situ} (single-site-resolved) and spin-resolved imaging technique \cite{Fukuhara1,Fukuhara2,Fukuhara3}, which is well-developed with a detection fidelity above $95\%$. In this way, one can obtain the Berry curvature from the relationship of Eq. (\ref{force}). The proposed system and procedure are schematically shown in Fig. 1: The atomic spin chain in the described evolution progress is initially prepared in the ground state and then is smoothly ramped along the angular direction, finally the magnetization and thus the Berry curvature is measured as a linear response to the ramp field. The quantization of the response will emerge in the Berry curvature of this interacting spin system, and in this sense, it is called the many-body dynamical QHE \cite{Gritsev2012}.

Now we turn to show that the dynamical QHE and the interaction-induced topological transition can be observed in the proposed system with numerical simulations. We numerically calculate the Berry curvature $F_{\theta\phi}$ from Eq. (\ref{force}) as a function of the spin-exchange interaction for the described ramp process by the time-dependent exact diagonalization \cite{ED}. The results for typical parameters and $N=8,9$ are shown in Fig. 2(a). As expected \cite{Gritsev2012}, for this spin chain system, the quantization plateaus in $F_{\theta\phi}$ appear in the value $n=0,1,2...,N/2$ ($n=1/2,3/2,...,N/2$) for even (odd) $N$ with decreasing the spin interaction. This phenomenon comes from the manifold of ground state degeneracies and the the Hilbert space topology in the parameter space. When the interaction $J$ tend to zero, the Berry curvature is equal to $N/2$ which indicates that the ground state behaves as a collective spin of magnitude $N/2$. When $J$ is large enough in the ferromagnetic regime, the system behaves as a spin singlet for even $N$ and as an effective spin $1/2$ for odd $N$. The transition between the spin singlet (for even chain) and the maximally polarized state occurs through the quantized steps, which reflect the total value of the spin in the initial state \cite{Gritsev2012}. From Fig. 2(a), we can find that all the quantized plateaus exhibit within the typical parameter regime of the interaction strength $J$ from 23 Hz to 280 Hz, with the fixed field strength $h=200$ Hz. In general, these plateaus in the dynamical response can be observed by varying the ratio $J/h$ (see Fig. 2(a) for the regime of ratio), such as changing $h$ with fixed $J$. Although our simulations are performed in a small interacting spin system, which are limited by the used time-dependent exact diagonalization method (that scales poorly
with system size) \cite{ED}, the characterizations of the many-body dynamical QHE (the quantization of the dynamical response and the interaction-induced topological transition) remain the same for a large system \cite{MC}.

To show that the quantization of the dynamical response holds under more generic conditions, we further consider the control errors which stem from the parameter fluctuations in the system Hamiltonian (\ref{SpinHam}). We assume the local fluctuations $\tilde{J}_j=\alpha_1 J$ and $\tilde{h}_j=\alpha_2 h$, with $\alpha_1$ and $\alpha_2$ randomly distributing in the region
$[1-\eta,1+\eta]$ for each lattice (spin) $j$ and $\eta>0$ as the fluctuation strength. For each single realization with randomly chosen $\alpha_{1}$ and $\alpha_{2}$, we repeat the similar ramp progress and calculate the corresponding $F_{\theta\phi}^\alpha$ as that in Fig. 2(a), and
then we can obtain the averaged Berry curvature $\bar{F}_{\theta\phi}=1/N_\alpha\sum
F_{\theta\phi}^\alpha$, where $N_\alpha$ denotes the sampling number. The averaged Berry curvature
for the system with $N=6$ as a function of the interaction strength is
plotted in Fig. 2(b), with $\eta=5\%,10\%,20\%$ and $N_{\alpha}=50$. One can find that the plateaus are still stable and the topological transition remains sharp, even when the parameter fluctuation strength up to $20\%$ and for a small sampling number. This result demonstrates that the quantized plateaus of the Berry curvature in the dynamical response are very robust against the control errors, making their observation in experiments more feasible. It is noted that in our simulations, the sharp topological transitions with the corresponding positions hold for different sampling numbers $N_{\alpha}$ (the simulation sizes), even for the single sampling case ($N_{\alpha}=1$) which is also shown in Ref. \cite{Gritsev2012}. However, with given fluctuation strengths, the plateaus become more stable and the quantization becomes more significant for larger $N_{\alpha}$.

\begin{figure}[tbp]
\includegraphics[width=18.0cm]{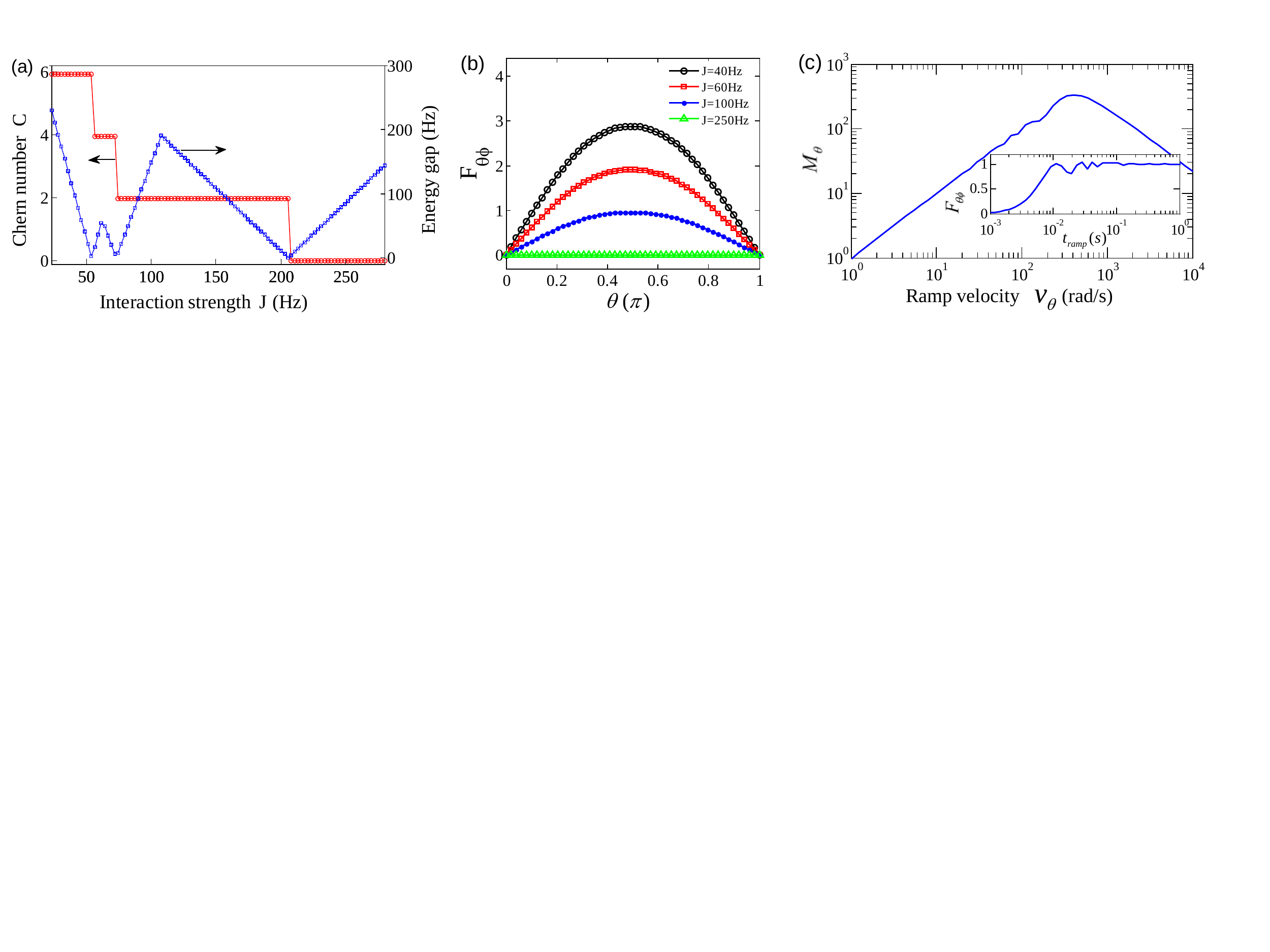}
\caption{(Color online) (a) The quantized Chern number and the energy gap between the ground state and the first excited state as a function of the interaction strength for the spin system with $N=6$ . (b) The Berry curvature $F_{\theta\phi}$ as a function of the finial mixing angle $\theta$ after ramp progressions, corresponding to the cases with typical parameters $J=40,60,100,250$ Hz in (a). (c) The magnetization (generalized force) $M_\theta$ as a function of the final ramp velocity $v_{\theta}$ for the spin system with $N=8$, showing the linear approximation within $v_{\theta}\lesssim
300$ rad/s. The inset shows the corresponding Berry curvature $F_{\theta\phi}$ as a function of the ramp time, which saturates to nearly one when $t_{\text{ramp}}\gtrsim 0.01$ s. Other parameters in (a) and (b) are $h=200$ Hz, and $J=h/2=100$ Hz in (c). }
\end{figure}

The quantization plateaus implies the topological property of the response. The corresponding topological index is the first Chern number $\mathcal{C}$, which can be obtained by integrating the Berry curvature over the whole spherical surface of the parameter space spanned by $\theta$ and $\phi$. For our case, the Hamiltonian is cylindrically invariant as one can get the Hamiltonian at arbitrary $\phi$ from the Hamiltonian at $\phi=0$ by just rotating the spins by an angle $\phi$ around the $z$-axis. Accordingly, the Berry curvature is cylindrically symmetric and is a function of $\theta$ alone. Therefore, one can obtain the Chern number as
 \begin{equation}
 \label{Ch}
\mathcal{C}=\frac{1}{2\pi}\int^{2\pi}_0d\phi\int^{\pi}_0d\theta F_{\theta\phi}=\int^{\pi}_0F_{\theta\phi}d\theta.
 \end{equation}
By simulating the $\theta$-ramp progressions from $\theta(t_{\text{ramp}})=0$ to $\pi$ via changing $t_{\text{ramp}}$, we calculate the corresponding Chern number from Eq. (\ref{Ch}). The results for the spin system with $N=6$ are shown in Fig. 3(a), where we also plot the energy gap between the ground state and the first excited state. As expected, the quantized Chern number is $\mathcal{C}=0,2,4,6$ as the interaction strength $J$ varies from $23$ Hz to $280$ Hz, and the energy gap closes at the topological transition points. Note that the Chern number is more robust and its quantized plateaus will be more significant due to the averaging over different runs of $\theta$ ramp with the parameter fluctuations \cite{Yang}. Figure 3(b) shows the Berry curvature as a function of the final mixing angle at the end of the ramp progressions, which correspond to the cases with the parameters $J=40,60,100,250$ Hz in Fig. 3(a). The sine profile in $F_{\theta\phi}(\theta)$ as shown in Fig. 3(b) is similar to that for a single spin in a driving magnetic field, and thus indicates the ground state of the system behaves as a collective spin.

\section{Discussions and conclusions}

Before concluding, we briefly discuss some possible concerns, the ramp velocity limit and the decoherence effect, in practical experiments. In our previous calculations, the Berry curvature $F_{\theta\phi}$ is considered to be a linear response to the ramp velocity $v_{\theta}$. In general cases, the magnetization (the generalized force) is determined by $M_\theta=M_0-F_{\theta\phi}
v_\theta+\mathcal{O}(v_\theta^2)$ \cite{Gritsev2012,Avron2011},
where the constant term $M_0$ gives the value of the magnetization
in the adiabatic limit and $M_0=0$ here. The linear
response approximation could break down when the velocity $v_{\theta}$ is too
large to neglect the term related to $v_\theta^2$. To check the
velocity limit for our proposed system in this linear response theory, we numerically
calculate the magnetization $M_\theta$ as a function of the finial ramp velocity $v_{\theta}$ for the system with $N=8$, as shown in Fig. 3(c). The result shows that the linear response approximation works well within $v_{\theta}\lesssim300$ rad/s. In addition, we calculate the corresponding Berry curvature as a function of the ramp time $F_{\theta\phi}(t_{\text{ramp}})$, which saturates to nearly one when $t_{\text{ramp}}\gtrsim 0.01$ s and becomes very stable when $t_{\text{ramp}}\gtrsim 0.06$ s. Therefore, to observe the quantized plateaus in Fig. 2(a), one requires the ramp velocity slower than about
300 rad/s, corresponding to the ramp time longer than about $0.01$ s. We also simulate the same procedures for the cases with $N=6,7,9$ and find that the results are similar to those in Fig. 3 (c), indicating the velocity limit for observing the quantized plateaus does not change much for systems of different sizes.

We note that the presented ultra-cold atomic spin chain should be quite robust to the realistic noise. First, since the atoms only virtually tunnel to the neighboring sites with a small probability about $(t/U)^2$, the dephasing rate and the inelastic decay rate are significantly reduced compared with the cold collision case \cite{BH,duan}, where the typical decoherence rate can be about several Hz \cite{BHExp}. In addition, the spontaneous emission noise rate can be made very small by using a blue-detuned optical lattice or by increasing the detuning. For example, using a blue-detuned lattice with a moderate detuning rate $5$ GHz, the effective spontaneous emission rate is estimated to be several Hz \cite{duan}. Under this condition, the noise-induced decoherence rate in our proposed cold atom system is significantly smaller (about two orders smaller) than the ramp rate, and thus the topological features of the dynamical QHE can be observed within the coherence time \cite{Yang}.

In conclusion, we have proposed a realistic scheme to simulate
the many-body dynamical QHE and the related interaction-induced topological
transition with ultra-cold atoms in an optical lattice. We have demonstrated that the
typical topological features of this effect can be observed under practical experimental conditions. The realization of the many-body dynamical QHE in the optical lattice may extend the current intensive study of topological states in cold atom systems and pave a new avenue to explore intriguing topological responses with interactions.

\begin{acknowledgments}
We thank Prof. S.-L. Zhu and Dr. Z.-Y. Xue for helpful discussions. This work was supported by the PCSIRT (Grant No. IRT1243) and the SRFYTSCNU
(Grant No. 15KJ16). X.-C. Yang was supported by grants from City University of Hong Kong (Project No. 9610335).

\end{acknowledgments}

\end{document}